\documentclass[twocolumn,showpacs,amsmath,amssymb]{revtex4}

\usepackage{graphicx}
\newcommand{\refb}[1]{(\ref{#1})}

\newcommand{\ZZ}{\mbox{\bf Z}}
\newcommand{\QQ}{\mbox{\bf Q}}
\newcommand{\Qp}{\mbox{\bf Q}_p}
\newcommand{\Zp}{\mbox{\bf Z}_p}

\begin{document}

%\preprint{hep-th/0606082}

\title{On the relation between $p$-adic and ordinary strings}

\author{Debashis Ghoshal}

\affiliation{Harish-Chandra Research Institute, Chhatnag Road,
Allahabad 211019, India.
{\tt ghoshal@mri.ernet.in}}

\begin{abstract}
 The amplitudes for the tree-level scattering of the open string
tachyons, generalised to the field of $p$-adic numbers, define the
$p$-adic string theory. There is empirical evidence of its relation to
the ordinary string theory in the $p\to 1$ limit. We revisit this
limit from a worldsheet perspective and argue that it is naturally
thought of as a continuum limit in the sense of the renormalization
group. 
\end{abstract}


\maketitle

%\section{Introduction}
The $p$-adic string theory was proposed\cite{FrOl} with a mathematical
motivation and with a hope that the amplitudes of these theories,
considered for all primes, will relate to those of the ordinary
strings through the adelic relation\cite{FrWi}. While this idea
remains to be realized, the early papers\cite{FrOl}--\cite{FrOka}
worked out the details of this theory. In particular, all the
tree-level tachyon amplitudes were computed from which the spacetime
effective theory of the tachyonic scalar was obatined and the
solutions of its equation of motion were studied. Subsequently a
`worldsheet' understanding was developed\cite{Spok}--\cite{CMZ} (see
\cite{BrFrRev} for a review).  More recently it has come into focus
through the realization that the exact spacetime theory of its tachyon
allows one to study the process of tachyon condensation. In
Ref.\cite{GhSeP}, the solitons of the effective theory of the $p$-adic
tachyon\cite{BFOW} were identified with the D-branes and shown that
the tachyon behaves according to the conjectures by Sen\cite{SenRev}.
(Henceforth we will refer to the $p$-adic string as $p$-string and its
tachyon as $p$-tachyon for brevity.)

An unexpected relation emerges with the ordinary bosonic open string
in the $p\to 1$ limit\cite{GeSh} (see also the prescient comments in
Ref.\cite{Spok}), when the effective action of the $p$-tachyon turns
out to approximate that obtained from the boundary string field
theory\cite{WiBSFT,ShBSFT} (BSFT) of ordinary strings. BSFT was useful
in proving the Sen conjectures\cite{GeSh,KuMaMo,GhSeN}. This correspondence
remains even after a noncommutative deformation of the $p$-tachyon
effective action.  In fact, thanks to it one can find {\it exact}
noncommutative solitons in BSFT (of the ordinary string theory) at
{\it all} values of the deformation parameter\cite{GhNCP}.

However, this relation in the $p\to 1$ limit is empirical. Moreover,
strictly $p$ can only take discrete values. In this Letter, we
consider the issue from a worldsheet point of view to advocate that
the limit is to be understood in terms of a sequence of string
theories based on (algebraic) extensions of increasing degree of the
$p$-adic number field $\Qp$. We argue that each of these provide a
discretization of the ordinary worldsheet (the disk or UHP) and their
effective actions relate to each other in terms of the renormalization
group (RG). There is a natural continuum limit in which the RG
transformed effective value of $p$ tends to one. A preliminary version
of these ideas was presented in the `12th Regional Conference on
Mathematical Physics' held in Islamabad, Pakistan\cite{GhIs}.

%We will refer to the $p$-adic string as $p$-string (and the tachyon of
%the $p$-adic string theory likewise as $p$-tachyon) for brevity. We
%flippantly suggest keeping the `$p$' silent so that it sounds the same
%as the `string'. More seriously, however, we will endeavour to show
%that in spite of the apparently different phenotypes of the $p$-string
%and the string, they share a closer genotypic relation.

%%%%%%%%%%%%%%%%%%%%%%%%%%%%%%%%%%%%%%%%%%%%%%%%%%%%%%%%%%%%%
%\section{The $p$-adic string: a recap}
Recall that the tree-level scattering amplitude of $N$ on-shell 
ordinary open-string tachyons of momenta $k_i$ ($i=1,\cdots,N$), $k^2_i=2, 
\sum k_i=0$ is given by the Koba-Nielsen formula in which the
integrals are over the real line {\bf R} and the integrand only
involves absolute values of real numbers:
\begin{equation}\label{Ntachampl}
{\cal A}_N = \!\!\int\! \prod_{i=4}^N d\xi_i\!
\left|\xi_i\right|^{k_1\cdot k_i}
\left|1-\xi_i\right|^{k_{2}\cdot k_i}\!
\!\!\!\!\!\prod_{4\le i< j\le N}\!\!\!\!\!\left|\xi_i-\xi_j
\right|^{k_i\cdot k_j}.
\end{equation}
Except for ${\cal A}_4$, the rest cannot be computed analytically.
Ref.\cite{FrOl} considered the above problem over the local field of
$p$-adic numbers $\Qp$, to which it admits a ready extension. In order
to describe it, let us digress briefly.

On the field of rational numbers {\bf Q}, the familiar norm is the
absolute value. The field {\bf R} of real numbers arise as the
completion of $\QQ$ when we put in the limit points of all Cauchy
sequences, in which convergence is decided by the absolute value norm.
However, it is possible to define other norms on {\bf Q} consistently.
To this end, fix a {\it prime} number $p$ and determine the highest
powers $n_1$ and $n_2$ of $p$ that divides respectively the numerator
$z_1$ and denominator $z_2$ in a rational number $z_1/z_2$, ($z_1,
z_2$ coprime).  The {\it $p$-adic norm} of $z_1/z_2$, defined as:
$\left|z_1/z_2\right|_p = p^{n_2-n_1}$, satisfies all the required
properties, indeed even a stronger version of the triangle inequality.
%\footnote{Norms with these properties are called
%non-archimedian.} $|x+y|_p\le\hbox{\rm max}(|x|_p,|y|_p)$. In fact,
%apart from the absolute value norm, the $p$-adic norms are the only
%possible ones (upto a natural notion of equivalence). 
The field $\hbox{\bf Q}_p$ is obatined by completing $\QQ$ using the 
$p$-adic norm. Any $p$-adic number $\xi\in\hbox{\bf Q}_p$ has a
representation as a Laurent-like series in $p$:
\begin{equation}\label{plaurent}
\xi = p^N\left(\xi_0 + \xi_1 p + \xi_2 p^2 + \cdots\right), 
\end{equation}
where, $N\in\ZZ$ is an integer, $\xi_n\in\{0,1,\cdots,p-1\}$,
$\xi_0\ne 0$ and $\left|\xi\right|_p=p^{-N}$. Deatils of materia
$p$-adica are available in {\it e.g.}, \cite{GGPSP}--\cite{GouP}; some
essential aspects are reviewed in \cite{BrFrRev}.

Coming back to the Koba-Nielsen amplitudes, Freund {\it et al}
modified these by replacing the absolute values by $p$-adic norms and
the real integrals by integrals over $\Qp$. These are, by definition,
the amplitudes for the scattering of $N$ open $p$-string tachyons. The
benefit is that all these integrals over $\Qp$ can be evaluated
analytically. Equivalently the tree level effective action of the open
$p$-string tachyon $T$ is known {\it exactly}\cite{BFOW,FrOka}: in
terms of a rescaled and shifted field $\varphi=1 + g_sT/p$:
\begin{equation}\label{paction}
{\cal L}_p\; =\;{p^2\over g^2(p-1)}\,\left[
-{1\over 2}\varphi\,p^{-{1\over 2}\Box}\varphi 
+ {1\over p+1}\,\varphi^{p+1}\right]. 
\end{equation} 
%%
%where $\varphi=1 + g_sT/p$ is a rescaled and shifted field.
%The $p$-tachyon potential has a local minimum and two (respectively
%one) local maxima for odd (respectively even) integer $p$.
%It also has pathological singularities at large
%values of the argument, as in the ordinary string. 

We emphasize that in the above, the boundary of the the open
$p$-string worldsheet is valued in $\Qp$, but the spacetime in which
the $p$-string propagates is the usual one. Once one arrives at the
spacetime action \refb{paction}, however, it can be extrapolated to
all integers.  Incidentally, there is also another extrapolation,
unrelated to this, in which the Veneziano amplitude (expressed in
terms of the gamma function) is modified to be valued in
$\Qp$\cite{Volo}--\cite{GhQAV}.

The equation of motion from \refb{paction} admits the constant
solutions $\varphi=1$ (unstable vacuum with the D-brane) and
$\varphi=0$. There is no perturbative open string excitation around
the latter, and hence is to be identified as the (meta-)stable closed
$p$-string vacuum. There are also soliton solutions. For any (spatial)
direction, there is a localized gaussian lump\cite{BFOW}.  When
identified as the different D-$m$-branes, the descent relations
between these confirm the Sen conjectures\cite{GhSeP}.

If one substitutes $p=1+\epsilon$ in \refb{paction} and takes the limit
$\epsilon\to 0$, one obtains\cite{GeSh},
%%
%\begin{equation}\label{bsftaction}
%{\cal L}_{p\to 1} =
%\half\varphi\,\Box\varphi + \half\varphi^2\left(\ln
%\varphi^2 - 1\right).  
%\end{equation}
%%
%This is, 
after a field redefinition $\varphi=e^{-T/2}$, the effective action of
the tachyon of the ordinary open string theory calculated from
BSFT\cite{WiBSFT,ShBSFT}. After a noncommutative deformation of
\refb{paction}, the gaussian soliton of $p$-string theory generalizes
to a one-parameter family of solitons, which are exact solutions to
the equation of motion. In the limit $p\to 1$, one finds an exact
solution to the ordinary string theory, where the noncommutativity
comes from a constant $B$-field background\cite{GhNCP}.
(Refs.\cite{GhKaBP,Grange} attempt to find the worldsheet origin of
the noncommutativity in $p$-string theory.)

%%%%%%%%%%%%%%%%%%%%%%%%%%%%%%%%%%%%%%%%%%%%%%%%%%%%
\begin{figure}[hb]
\begin{center}
  \includegraphics[width=4.0cm]{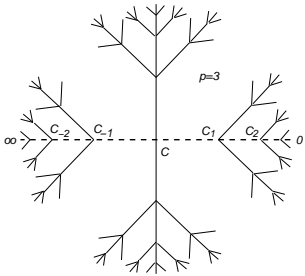} \caption{\label{tree} 
  The `worldsheet' of the $3$-adic string ${\cal B}_3$, 
  $\partial{\cal B}_3=\hbox{\bf Q}_3$. The dotted line is the 
  path from the boundary points 0 to $\infty$.}
\end{center}
\end{figure}

%%%%%%%%%%%%%%%%%%%%%%%%%%%%%%%%%%%%%%%%%%%%%%%%%%%%
%\section{`Worldsheet' of the $p$-adic string}
At first sight the relation to the ordinary strings is all the more
surprising and counter-intuitive from the point of view of the
$p$-string `worldsheet'. In fact, the `worldsheet' itself, the
boundary of which is $\Qp$, is not in the least
obvious\cite{ZabroWS,CMZ}. At tree level, the analog of the unit disk
or the UHP of the usual theory, is an infinite lattice with no closed
loops, {\it i.e.}, a uniform tree ${\cal B}_p$ in which $p+1$ edges
meet at each vertex (see Fig.~\ref{tree}). This is the familiar Bethe
lattice ${\cal B}_p$, known in the context of $\Qp$ as the Bruhat-Tits
tree.  Its boundary, defined as the union of all infinitely remote
vertices, can be identified with $\Qp$. To see this, one may use {\it
  e.g.}, the representation \refb{plaurent} in which case, the integer
$N$ chooses a branch along the dotted path (in Fig.~\ref{tree}) and
the infinite set of coefficients $\xi_n$ determine the path to the
boundary. On the other hand, the tree ${\cal B}_p$ is the (discrete)
homogeneous space PGL(2,$\Qp$)/PGL(2,$\Zp$).
%: the coset obtained by modding PGL(2,$\Qp$)
%by its maximal compact subgroup PGL(2,$\Zp$). 
This construction parallels the case of the ordinary string theory.
%, in which the UHP is
%the homogeneous coset PSL(2,$\hbox{\bf R}$) modulo its maximal compact
%subgroup SO(2).  The action of PGL(2,$\Qp$) on $\Qp$ extends naturally
%to ${\cal B}_p$.

The Polyakov action on the `worldsheet' ${\cal B}_p$ is the natural
discrete lattice action for the free massless fields $X^\mu$. The
action of the laplacian at a site $z\in{\cal B}_p$ is
%%
%\begin{equation}\label{treelaplace}
$\nabla^2 X^\mu(z) = \sum_{i}  X^\mu(z_i) - (p+1) X^\mu(z)$, 
%\end{equation}
%%
where, $z_i$ are the $p+1$ nearest neighbors of $z$. It was shown in
\cite{ZabroWS} that starting with a finite Bethe lattice and inserting the 
tachyon vertex operators on the boundary, one recovers the
prescription of \cite{FrOl,BFOW} in the thermodynamic limit.  

Naively the lattice is one dimensional for $p=1$. However, the
relation to the ordinary string is through the limit $p\to 1$ and it
is not apparent how to make sense of this for the discrete variable
$p$. This is the problem we will address in the following. First, we
claim that ${\cal B}_p$ gives a discretization of the disk/UHP. This
does not seem possible because in ${\cal B}_p$, the number of sites
upto some generation $n$ from an origin $C$ (say) grows exponentially
for large $n$:
\begin{equation}\label{growth}
{\cal N}_n %= 1 + (p+1) + (p+1)p + \cdots + (p+1)p^n
\sim \exp(n\,\ln p).  
\end{equation}
Therefore, its formal dimension is infinite. Indeed, Bethe lattices
are used in calculating the results in the upper critical dimension of
model theories. For example, for a free scalar field theory with
arbitrary interactions (from, say, vertex operators) the upper
critical dimension is two. One would expect to get this from a Bethe
lattice.

The tacit assumption above is that the embedding is in an {\it
  Euclidean} space.  On the other hand, in a $d$-dimensional {\it
  hyperbolic} space with the metric $ds_H^2 = dr^2 + R_0^2
\sinh^2\left({r\over R_0}\right)\, d\Omega_{d-1}^2$ the volume of a
ball of radius $R$ ($R>>R_0$, the radius of curvature) also grows
exponentially for large $R$:
\begin{equation}\label{hgrowth}
\hbox{\rm vol}_d(R) \sim \exp\left({d-1\over R_0}\,R\right). 
\end{equation}
This suggests a natural embedding of ${\cal B}_p$ in hyperbolic
spaces. Parametrizing
\begin{equation}\label{pparam}
p = 1 + {a\over R_0}(d-1),  
\end{equation}
and considering the limit $a\to 0$ so that $p\to 1$, the formulas
\refb{growth} and \refb{hgrowth} agree for
$\displaystyle\lim_{{n\to\infty\atop a\to 0}}n\,a = R$, from which $a$
is seen as the lattice spacing. Thus a uniform Bethe lattice ${\cal
  B}_p$ can be used to discretize a hyperbolic space of constant
negative curvature. Moreover, $p\to 1$ provides a natural continuum
limit. This is true, in particular, when the dimension $d=2$, the case
of our interest. In fact the embedding of ${\cal B}_p$ into the unit
disk/UHP equipped with, say, the Poincar\' e metric), is {\it
  isometric}. It is related to the hyperbolic tessellation of the
disk/UHP and often has interesting connection with the fundamental
domains of the modular functions of SL(2,{\bf C}) and its
subgroups\cite{htess}.

%%%%%%%%%%%%%%%%%%%%%%%%%%%%%%%%%%%%%%%%%%%%%
\begin{figure}[hb]
\begin{center}
  \includegraphics[width=4.0cm]{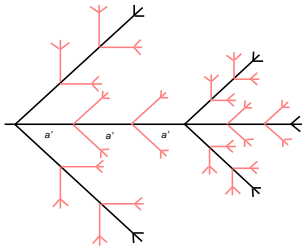} \caption{\label{finebethe} 
  The lattice with spacing $a'$ leads to a coarse grained one with spacing 
  $a=ma'$ ($m=3$ here), when the `grey' branches are integraded out.}
\end{center}
\end{figure}
%%%%%%%%%%%%%%%%%%%%%%%%%%%%%%%%%%%%%%%%%%%%%

The standard way to obtain a continuum limit from a lattice
regularization is to go to lattices with smaller lattice spacings and
eventually consider the limit in which this becomes vanishingly small.
%In Fig.~\ref{finebethe}, the `black'
%lattice with lattice spacing $a=3a'$ gives a coarser approximation
%compared to that with spacing $a'$. 
Suppose we start with the `black' sublattice in Fig.~\ref{finebethe}, the 
boundary of which is $\mbox{\bf Q}_3$. In comparing this to 
the full lattice, we see that between two
neighbouring `black' nodes there are two `grey' nodes, which in turn
branch further so that the full lattice is similar to the `black'
one. What, if any, is the relation of the full lattice to $\mbox{\bf
Q}_3$? To answer this question, we need to recall some facts
about $\Qp$.

%%%%%%%%%%%%%%%%%%%%%%%%%%%%%%%%%%%%%%%%%%%%%%%%%%%
%\section{A digression to the extensions of the $p$-adic number field}
The field $\Qp$ (like {\bf R}) is not closed algebraically.  That is,
not all roots of polynomials with coefficients in the field belong to
it. For {\bf R}, one can adjoin a root of $x^2+1=0$ and extend to the
algebraically closed (and complete) field of complex numbers {\bf C}.
It is said to be an {\it index} two extension, {\it i.e.}, {\bf C} is
a two dimensional vector space over {\bf R}.  The story is more
complex for $\Qp$, for which there are infinitely many algebraic
extensions none of which is closed.
%In fact, none of
%the finite algebraic extensions of $\mbox{\bf Q}_p$ is closed. To get
%a closed field one needs to consider the union of all such
%extensions\footnote{It turns out that it is not complete. Thankfully,
%after completion, the resulting field is still closed and is analogous
%to $\CC$ in a sense. We will not consider it any further, although
%this may turn out to be the right setting for the {\it closed}
%$p$-strings.}.  
Now consider a finite extension $\overline{\mbox{\bf Q}}_p^{(n)}$ of
index $n$.  There are several such and an integer $e$, called the {\it
  ramification index} partially distinguishes between them. It turns
out that $e$ divides $n$, so that $f=n/e$ is again an
integer\cite{GGPSP}--\cite{GouP}. First, we will consider a so called
{\it totally ramified} extension for which $e=n$.  The Bruhat-Tits
tree for such extensions can be obtained from the original one of
$\Qp$ through the process described in the last paragraph.  Namely, to
get the tree for a totally ramified extension
${\overline{\QQ}}_p^{(n)}$, start with the `black' tree for $\Qp$ and
introduce $(e-1)$ new nodes between the exisiting ones. Connect
(infinite) `grey' branches to these so that the tree is uniform with
coordination number $p$ as before. In the other cases when $e<n$, one
also needs to introduce an infinite number of new edges and nodes so
that the resulting tree is uniform with coordination number
$p^f$\cite{ZabroWS,CMZ}.

In $\overline{\QQ}_p^{(n)}$, there is a special element $\pi$, called
the {\it uniformizer}, that plays the role of $p$ for $\Qp$.
Specifically, any element of $\overline{\QQ}_p^{(n)}$ can be expressed
as a Laurent series in terms of $\pi$ (just like \refb{plaurent}), and
the norms of its elements are integer powers of $\pi$. In particular,
for $p\in\overline{\QQ}_p^{(n)}$:
\begin{equation}\label{ppi}
p\simeq\pi^e, 
\end{equation}
where the approximate equality indicates the leading term in the
expansion. Parametrizing both $p$ and $\pi$ as in \refb{pparam}, $a'$
of $\overline{\cal B}_p^{(n)}$ is related to $a$ of ${\cal B}_p$ as
$a\simeq na'$, as is apparent from the construction. Thus for larger
and larger extensions $\pi\simeq p^{1/e}$ approaches the value 1 for
any $p$. The corresponding lattices provide finer discretizations and
a passage to the continuum limit.

%%%%%%%%%%%%%%%%%%%%%%%%%%%%%%%%%%%%%%%%%%%%%%%%%%%%%%
%\section{Sequence of non-archimedian strings and the renormalisation group} 
The construction of the previous section suggests a way to understand
the limit $p\to 1$ through a sequence of string theories based on the
extensions of $\Qp$. For simplicity let us consider a totally ramified
extension. Apparently there is a puzzle. The tachyon amplitudes for
the totally ramified extension $\overline{\QQ}_p^{(e=n)}$, turn out to
be exactly the same as those for $\Qp$! This is because the {\it
  coefficients} in the Laurent expansions of both are from the same
set; the trees are similar, therefore, the measures that affect the
integrals work out to be identical\cite{BFOW}. Hence, the effective
action of the tachyon of these two theories are identical.  String
theories based on extensions of $\Qp$ were already considered in
\cite{FrOl,BFOW}, indeed the very first paper on $p$-adic string
theory\cite{FrOl} dealt with the quadratic extensions of $\Qp$. In
analogy with ordinary strings, it was thought to be a theory of closed
strings. The theories based on higher extensions were called {\it even
  more closed} strings! In hindsight, it is natural to think of all
these as open strings.

Returning to the apparent paradox, the resolution comes from the
following. In taking a continuum limit, one is not really interested
in the results separately for the two theories, but rather in
comparing the degrees of freedom of the coarse-grained lattice from
the fine one from the perspective of a (real space) RG. In order to do
this, only the degrees of freedom on the `grey' nodes and branches
(see Fig.~\ref{finebethe}) should be integrate out. This leaves one
with the `black' sublattice with some effective interaction between
these residual degrees of freedom. A rescaling of the lattice so that
the spacing $a\to ba=a'$ completes the RG transformation.

Let us see the effect of these on the Poisson kernel on the Bethe
lattice. It is more transparent for the Dirichlet problem for which
the Green's function is\cite{ZabroWS}
\begin{equation}\label{dirichlet}
{\cal D}(z,w)={p\over p^2-1} p^{-d(z,w)}, 
\end{equation}
where $d(z,w)$ is the number of steps in lattice units between the
sites $z,w$. Since the spacing in ${\cal B}_p$ is $e=n$ times that in
$\overline{\cal B}_p^{(n)}$, $d_{{\cal B}} = e d_{\overline{\cal
    B}^{(n)}}\equiv e\overline{d}$ and after integrating out the
intermediate sites, ${\cal D}_{\hbox{\rm\small eff}}={p\over p^2-1}
p^{-e\overline{d}(z,w)}$. When the lattice is rescaled, the original
form of the kernel is recovered with the substitution
$p\to\pi=p^{1/e}$. The Green's function ${\cal N}(z,w)$ for the
Neumann problem is roughly the logarithm of ${\cal
  D}(z,w)$\cite{ZabroWS}, so the same argument holds there as well.
Thus the effect of the RG transformation on the tachyon action
\refb{paction} is to replace $p\to\pi=p^{1/e}$. The action for the
ordinary bosonic string is obtained in the limit $e\to\infty$, which
is a continuum limit in the sense of RG.

The above argument can be straightforwardly extended to any finite
extension of $\Qp$. 
%The tachyon amplitudes for such an extension are
%obtained from those of $\Qp$ by the substitution $p\to p^f$\refb{BFOW},
%hence, the same holds for the effective action. The net effect of the
%two step RG transformation is to replace $p\to p^{1/n}$ in the
%effective action. For extensions of very large degree, {\it i.e.}, for
%$n\to\infty$, we have $p_{\hbox{\rm\small eff}}\to 1$ and we get the action
%\refb{bsftaction}, which is an approximation to that obtained from BSFT.
%
Let us also note that only the unramified extension ($e=1$) is unique;
there are several partially and totally ramified extensions differing
in the details of the structure of the field. However, the associated
Bruhat-Tits trees, which are the objects of interest to us, are
specified only by the values of $e$ and $f$. It is not clear to us if
the non-uniqueness has any role to play for the string theories based
on these fields.

Further evidence comes from the problem of a random walk on a Bethe
lattice, for which Ref.\cite{MoTe} found an exact solution. This goes
over to the solution of the Brownian motion on a hyperbolic space of
constant negative curvature in the (formal) limit $p\to 1$. Thus the
Green's function for the diffusion equation on the hyperbolic disk/UHP
can be obtained as a continuum limit from the Bethe lattice.  The well
known relation between the kernel of the diffusion equation and the
Green's function of a free scalar field theory, can be to used obtain
the latter.  We are interested in a diffeomorphism and Weyl invariant
free scalar field theory coupled to the metric on the disk/UHP. There
are also marked points corresponding to asymptotic states given by
vertex operators on its boundary. Only hyperbolic metrics can be
consistently defined on such a surface. Further, with the freedom from
diffeomorphism and Weyl invariance, the metric can be made one of
constant negative curvature. In the worldsheet functional integral,
therefore, the contribution is from such a surface. The continuum
limit of a scalar field theory on a Bethe lattice would seem to give a
good approximation.

%%%%%%%%%%%%%%%%%%%%%%%%%%%%%%%%%%%%%%%%%%%%%%%%%%%%
%\section{Summary and some comments}
In summary, we have argued that the observation that the effective
field theory of the tachyon of the $p$-adic string approximates that
of the ordinary string in the $p\to 1$ limit, can be understood in
terms of RG flow on a sequence of open string theories based on
(algebraic) extensions of increasing degree of the $p$-adic field.
Each of these theories provides a discretization of the tree-level
worldsheet of the ordinary string and the $p\to 1$ limit is a
continuum limit in the sense of (real space) RG.

A few brief closing remarks. First, in the $p$-adic discretization,
the `worldsheet' is isometric to the disk/UHP with a metric of
constant negative curvature. This is a solution to the equation of
motion of Liouville field theory, and is interpreted as the
D0-brane\cite{ZaZa}.  Secondly, there is a more standard
discretization in terms of large $N$ random matrices. The zeroes of
the partition function of of the Ising and Potts models on random
lattices from $1\times1$ matrices and on Bethe lattices are
identical\cite{ThinF}, suggesting some kind of complimentarity in the
two discretizations.  Finally, the $p\to 1$ limit in terms of a set of
theories based on extensions of $\Qp$ may be useful in finding the
`closed' strings of the $p$-adic theory.

\smallskip

%%%%%%%%%%%%%%%%%%%%%%%%%%%%%%%%%%%%%%%%%%%%%%
%\section*{Acknowledgments}   
%\noindent{\bf Acknowledgments:}
It is a pleasure to thank Chandan Dalawat, Peter Freund and Stefan
Theisen for useful discussions. Hospitality at the Albert Einstein
Institute, Germany, where a part of the work was done, is acknowledged
gratefully.

%%%%%%%%%%%%%%%%%%%%%%%%%%%%%%%%%%%%%%%%%%%%%%

\end{document}